\documentstyle[aps,prb,twocolumn]{revtex}
  \title{Electron-hole correlations in semiconductor quantum dots with
tight-binding wave functions }
  \author{Seungwon~Lee, Lars~J\"onsson, and John~W.~Wilkins}
  \address{Department of Physics, Ohio State University, Columbus, Ohio 43210-1106}
  \author{Garnett~W.~Bryant}
  \address{National Institute of Standards and Technology, Gaithersburg,
Maryland 20899-8423}
  \author{Gerhard~Klimeck}
  \address{Jet Propulsion Laboratory, California Institute of Technology,
Pasadena, California 91109}
  \date{\today}

\newcommand{\ba}{\begin{eqnarray}}
\newcommand{\ea}{\end{eqnarray}}
\newcommand{\der}{{\rm d}}

\begin{document} 
\twocolumn[\hsize\textwidth\columnwidth\hsize\csname
@twocolumnfalse\endcsname 
\maketitle 
\widetext 
\begin{abstract} 
The electron-hole states of semiconductor quantum dots are investigated within the
framework of empirical tight-binding descriptions for Si, as an example of an
indirect gap material, and InAs and CdSe as examples of typical III-V and
II-VI direct-gap materials. The electron-hole Coulomb interaction is 
 largely insensitive to both the real-space description of the atomic basis
orbitals and different ways of optimizing the tight-binding parameters.  
Tight-binding parameters that are optimized to give the best
effective masses significantly improve the energies of the excitonic
states due to the better single-particle energies. However,
the Coulomb interaction does not vary much between different parameterizations.
In addition, the sensitivity of the Coulomb interaction to the choice of atomic orbitals
decreases with increasing dot size. Quantitatively, tight-binding
treatments of correlation effects are reliable for dots with radii 
larger than 15--20 \AA.
The calculated excitonic gaps are in good agreement with
recent photoluminescence
data for Si and CdSe but agree less well for InAs. 
Further, the effective range of the
electron-hole exchange interaction is investigated in detail.
In quantum dots of the direct-gap
materials InAs and CdSe, the exchange interaction can be long-ranged, extending
over the whole dot when there is no local (onsite) orthogonality between
the electron and hole wave functions.  By contrast, for Si quantum dots the
extra phase factor due to the indirect gap effectively limits the range to
about 15 \AA\@, independent of the dot size.  
\end{abstract} 
\bigskip \bigskip] \narrowtext 

\section{Introduction} \label{sec:intro}

  The optical and electronic properties of semiconductor quantum dots have
been studied both experimentally\cite{calcott}$^-$\cite{millo} and
theoretically\cite{efros}$^{-}$\cite{delerue} for a wide range of sizes, shapes
and materials.  This work is stimulated both by a fundamental interest in
quantum-confined systems and by the applicability of quantum dots
in nanoscale devices. Experimentally, significant recent improvements in both
growth techniques\cite{growthtech} and single-dot spectroscopy
\cite{singledotspec} have enabled
detailed
studies of the energy spectra of electron-hole complexes (excitons,
biexcitons, trions, etc.). Theoretically, the most sophisticated theoretical
approaches are multi-band effective-mass theory,\cite{banin}$^{,}$\cite{efros}
empirical pseudo-potential,\cite{ogut}$^{-}$\cite{williamson} tight-binding
methods,\cite{delerue2}$^{-}$\cite{allan} and quasiparticle calculations\cite{delerue} with the {\it GW} approximation.\cite{GW}

Quantum dots are intermediate between molecular and bulk systems. This is
reflected in the different theoretical approaches; effective mass theory
treats the dot as a confined bulk system whereas pseudo-potential theory aims
at a detailed atomistic description of the wave functions.  Tight-binding
theory compromises between these two approaches by including an atomistic
description but limiting the local degrees of freedom to a small basis set.  
Therefore, the computationally less costly tight-binding method can be used to
study large quantum dots, up to 25~nm size, without severely restricting
atomic-scale variations in the wave functions.  However, since the tight-binding
matrix elements are empirically optimized without introducing any specific
atomic orbitals, there is no direct way to calculate other matrix elements such as  Coulomb and exchange matrix elements.  Therefore, calculations involving
electron-hole interactions require a selection of specific atomic orbitals
that cannot be explicitly related to the tight-binding parameters.  Hence, a key
question that we address in this work is  how sensitive the calculated
electron-hole properties are to specific choices of orbitals used to
calculate the Coulomb matrix elements.

  We study spherical semiconductor crystallites centered around
an anion atom in a zincblende structure. We choose Si, InAs, and CdSe 
 as examples of an indirect-gap material, and typical III-V and II-VI
direct-gap materials respectively.  Experimentally, the low-lying exciton
energies have been measured for dots up to 20 \AA\@ radius in Si,\cite{wolkin} and up to 40
\AA\@ radius in InAs\cite{banin} and CdSe.\cite{norris} 

We use the empirical nearest-neighbor
$sp^3s^*$ tight-binding model\cite{vogl} for the electron and hole
single-particle wave functions.
In order to calculate electron-hole Coulomb and exchange matrix elements, we describe the real-space atomic basis orbitals
 $s,p_x,p_y,p_z,$ and $s^*$
with Slater orbitals as a starting point.
 Both the Coulomb and exchange interaction are screened by a
dielectric function depending on both dot size and the distance between the
particles.  The energies of the electron-hole states are obtained by
diagonalizing the configuration interaction matrix generated by the lowest-lying
electron and hole states with typical convergence of a few meV.  We
examine the sensitivity of the electron-hole energies on both the choices of
 atomic basis orbitals and the tight-binding parameters.

Within the tight-binding model, the 
single-particle Hamiltonian can be improved 
by either increasing the number of
basis orbitals or including interactions between more distant atoms. 
An $s^*$
orbital was first introduced  by Vogl {\it et al.}\cite{vogl} to improve the conduction 
bands near the $X$ point. 
To some degree, an $s^*$ orbital can mimic band structure effects that should
be attributed to $d$ bands. However, it is likely that $d$ orbitals need to
be added in some materials.\cite{allan} An alternative is to add
next-nearest neighbor interactions\cite{delerue2} within the $sp^3s^*$ model, which could
improve the band structure without increasing the computational cost of
generating Coulomb matrix elements, since the number of orbitals is
unchanged. Nevertheless, for the topics we discuss in this work the
nearest-neighbor $sp^3s^*$ model is a good starting point.
We here focus on how the single-particle energies can be improved by
tight-binding parameters specifically optimized to give good effective masses,
and on  how single-particle wave functions from different  
parameter sets affect the two-particle Coulomb interactions.

One interesting issue concerning quantum dots is 
the range of the electron-hole exchange interaction.
Franceschetti {\it et al.}\cite{franceschetti_ex} 
show that due to the lack of local orthogonality
between electron and hole wave functions  the exchange interaction can extend over the
whole dot.
We investigate in detail the effective range of the exchange interaction by
applying a cutoff range to the Coulomb potential. The origin of the
characteristic range of the exchange interaction is revealed by the analysis of 
the `exchange charge density' of the electron-hole pair. 

\section{Theory} \label{sec:theory}

\subsection{Hamiltonian of an electron-hole pair} \label{subsec:hamiltonian}

  The effective Hamiltonian of an electron-hole pair contains a
single-particle term and an electron-hole term
as follows (excluding spin-orbit coupling):
  \ba
  H &=& \sum_s \int \der^3 r ~ \hat{\psi}^{+}_{s} ({\bf r}) ~ \{ -
\frac{1}{2} \nabla^2 + V({\bf r}) \} ~ \hat{\psi}_{s}({\bf r}) \nonumber\\
  &&+\frac{1}{2}\sum_{s',s}\int\!\!\int \der^3 r' ~ \der^3  r ~
\frac{\hat{\psi}^{+}_{s'} ({\bf r'}) \hat{\psi}^{+}_{s} ({\bf r}) \hat{\psi}_{s}({\bf r})
\hat{\psi}_{s'}({\bf r'})}{\epsilon(|{\bf r'}-{\bf r}|,R) ~ |{\bf r'}-{\bf r}|},
  \label{eq:total_H}
  \ea
  where
  \ba
  \hat{\psi}_s({\bf r}) &=& \sum_n \hat{a}_{n,s} \phi_{n}({\bf r}) + \hat{b}^+_{n,-s} \phi^*_{n}({\bf r}), \nonumber \\
  \hat{\psi}_s^+({\bf r}) &=& \sum_n \hat{a}^+_{n,s} \phi^*_n({\bf r}) + \hat{b}_{n,-s} \phi_{n}({\bf r}),
  \ea
  with  the spin index $s$ and the tight-binding orbital index $n$, which
includes  atom-site index $i$ and  orbital-type index $\gamma$. 
 The functions $\phi_{n}({\bf r})$ are tight-binding basis
orbitals, which in this work are $s, p_x, p_y, p_z$ or $s^*$.
 The operators $\hat{a}_{n,s}$, $\hat{a}^+_{n,s}$ and $\hat{b}_{n,s}$, $\hat{b}^+_{n,s}$
 are annihilation and creation operators of an electron and a hole 
for the tight-binding basis orbital $\phi_{n}({\bf r})$, respectively.
Note that $a_{n,s}$ and $a_{n,s}^+$ are applied to conduction electron states, while $b_{n,s}$ and $b_{n,s}^+$ are applied to valence electron states.  
The dielectric function $\epsilon(|{\bf r'}-{\bf r}|,R)$ is assumed to be 
a function of
both the separation $|{\bf r'}-{\bf r}|$ of two particles and the dot radius
$R$. Note that we use atomic units for all the equations in this paper. 

The single-particle potential $V$ is implicitly defined through the
empirical tight-binding matrix elements. We use the nearest-neighbor $sp^3s^*$ 
tight-binding description.
The structure of a quantum dot is modeled as an anion-centered zincblende 
structure.\cite{note_cdse} 
Dangling bonds on
the surface are removed by explicitly shifting the energies of the corresponding hybrids well
above the highest calculated electron states.
This treatment mimics a dot whose surface is 
efficiently passivated with, for example, hydrogen or ligand molecules.
 
By multiplying the electron and hole eigenstates from the solution of the
tight-binding Hamiltonian, we obtain an electron-hole basis set
$|eh\rangle|j_s,m_s\rangle$ where $|j_s, m_s\rangle$ is an electron-hole spin state, i.e., either the singlet component $|0,0\rangle$ or one of the triplet components $|1,1\rangle, |1,0\rangle$, or $|1,-1\rangle$.
The spatial part $|eh\rangle$ is the product of an electron and a hole wave function that in real
space has the form:
\ba
 \psi_e({\bf r}_e)\psi_h^*({\bf r}_h) &=& \sum_{n_1,n_2} c_{e;n_1} c_{h;n_2}^* \phi_{n_1}({\bf r}_e)\phi_{n_2}^*({\bf r}_h).
\ea
This description closely follows that of Leung and Whaley.\cite{leung_si}

  The single-particle Hamiltonian  can be written in terms
of the electron-hole basis set and its eigenvalues:
  \ba
  H_{\rm{single}} &=& \sum_{eh j_s m_s} (E_e-E_h) |eh \rangle |j_s m_s\rangle \langle j_s m_s| \langle eh|,\label{eq:single}
  \ea
  where $E_e$ and $E_h$ are the electron and hole energies of the single-particle Hamiltonian.

  Projecting the electron-hole Hamiltonian onto the two-particle basis
set yields the electron-hole Hamiltonian with a Coulomb interaction
$J$ and an exchange interaction $K$:\cite{leung_si}$^,$\cite{knox}
  \ba
  H_{\rm e-h} &=& \sum_{j_s m_s} (J + K) |j_s m_s\rangle \langle j_s m_s|,
  \ea
  \ba
  J &=& -\sum_{e'h'eh} |e'h' \rangle \langle eh| \nonumber\\
  &&\times \int\!\!\int \der^3 r' \der^3 r\frac{\psi^*_{e'}({\bf r'})\psi_{e}({\bf r'}) \psi^*_{h}({\bf r})
\psi_{h'}({\bf r})}{\epsilon(|{\bf r'}-{\bf r}|,R)~|{\bf r'}-{\bf r}|}, \\
  K &=& 2\delta_{j_s} \sum_{e'h'eh} |e'h'\rangle \langle eh| \nonumber\\
  &&\times \int\!\!\int \der^3 r' \der^3 r\frac{\psi^*_{e'}({\bf r'})\psi_{h'}({\bf r'})
\psi^*_{h}({\bf r})\psi_{e}({\bf r})}{\epsilon(|{\bf r'}-{\bf r}|,R)~|{\bf
r'}-{\bf r}|} \label{eq:exch_eh},
\ea
   where $\delta_{j_s}$ is
unity for a singlet state 
 and zero for a triplet state. 
 The factor 2 in front of $\delta_{j_s}$ in Eq.~(\ref{eq:exch_eh}) is due to the fact that the exchange interaction allows two final electron-hole spin states $|\uparrow_e,\downarrow_h \rangle$ and $|\downarrow_e,\uparrow_h\rangle$ for an initial state $|\uparrow_e,\downarrow_h \rangle$ (or $|\downarrow_e,\uparrow_h \rangle$). In contrast, the Coulomb interaction requires that the spin of the electron should be the same between an initial and a final state, as should the spin of the hole. 

The Coulomb interaction $J$ describes the scattering of the
electron from $e$ to $e'$ and the hole from $h$ to $h'$, whereas the exchange
interaction $K$ describes the recombination of a pair $e$,$h$ at ${\bf r}$ and the
recreation of a pair $e'$,$h'$ at ${\bf r'}$.
Since we do not include spin-orbit couplings, the spin states are not coupled to one another in the present Hamiltonian. Therefore, we will use only the spatial part $|eh\rangle$ of the electron-hole basis set from now on.
The only constraint that the spin state gives to the Hamiltonian is the spin-selection rule in $K$.

The matrix elements of the electron-hole interaction Hamiltonian can now be rewritten in terms of 
integrals over the tight-binding basis orbitals by replacing $\psi_e({\bf r}_e)$ with $\sum_{n}c_{e;n}\phi_n({\bf r}_e)$ and $\psi_h({\bf r}_h)$ with $\sum_n c_{h;n}\phi_n({\bf r}_h)$:
  \ba
  \lefteqn{\langle e'h'|J|eh\rangle} \nonumber\\
  &=& - \sum_{\{n\}} c^*_{e';n_1} c_{e;n_2} c^*_{h;n_3} c_{h';n_4} ~
\omega(n_1,n_2;n_3,n_4),\\
  \lefteqn{\langle e'h'|K|eh\rangle} \nonumber \\
  &=& 2\delta_{j_s} \sum_{\{n\}} c^*_{e';n_1} c_{h';n_2} c^*_{h;n_3} c_{e;n_4}
~ \omega(n_1,n_2;n_3,n_4), \label{eq:h_eh}
  \ea
  where
  \ba
  \lefteqn{\omega(n_1,n_2;n_3,n_4)} \nonumber \\
  &=& \int\!\!\int \der^3 r'  ~ \der^3 r
\frac{\phi^*_{n_1}({\bf r'}) \phi_{n_2}({\bf r'})
\phi^*_{n_3}({\bf r}) \phi_{n_4}({\bf r})} {\epsilon(|{\bf r'}-{\bf r}|,R) ~
|{\bf r'}-{\bf r}|}.\label{eq:omega4}
  \ea

 Following Leung and Whaley,\cite{leung_si}  we approximate 
 the Coulomb and exchange interaction matrix elements 
by considering only terms having at most two distinct basis orbitals.
This approximation is reasonable since the integrals involving more than 
two different orbitals are typically small compared to the kept
integrals.\cite{martin}$^,$\cite{leung_si}
Integrals in Eq.~(\ref{eq:omega4}) with $n_1=n_2$ and $n_3=n_4$ are \emph {Coulomb
integrals} and those with $n_1=n_4$ and $n_2=n_3$ or with $n_1=n_3$ and $n_2=n_4$
 are \emph {exchange integrals}. Hence,
we define:
  \ba
  \omega_{\rm Coul}(n_1,n_2) &\equiv& \omega(n_1,n_1;n_2,n_2) \label{eq:Coul}\\
  \omega_{\rm exch}(n_1,n_2) &\equiv& \omega(n_1,n_2;n_2,n_1) \nonumber \\
                             &=& \omega(n_1,n_2;n_1,n_2)  \label{eq:exch}, 
  \ea
  where the equality of the two exchange integrals is valid when the tight-binding orbitals are real, as they are in this work.

  To make our notations clear, note that the Coulomb and the exchange \emph{integrals} of the basis
orbitals are the interactions between the tight-binding basis orbitals. By
contrast, the Coulomb and the exchange \emph{interactions} $J$ and $K$ are
interactions between the electron and hole wave functions. In fact, the
Coulomb interaction has contributions from both the Coulomb
and exchange integrals as does the exchange interaction.

  Finally, we can describe the electron-hole matrix elements in
terms of the Coulomb and exchange integrals.
  \ba
  \lefteqn{\langle e'h'|J|eh \rangle}\nonumber \\
  &=& -\sum_{n_1,n_2} c^*_{e';n_1} c_{e;n_1}
c^*_{h;n_2} c_{h';n_2} ~ \omega_{\rm Coul} (n_1,n_2) \nonumber \\
  &&- \sum_{n_1,n_2} c^*_{e';n_1}  c_{e;n_2} c^*_{h;n_1} 
c_{h';n_2} ~ \omega_{\rm exch} (n_1,n_2) \nonumber \\
  &&- \sum_{n_1,n_2} c^*_{e';n_1}  c_{e;n_2} c^*_{h;n_2}
c_{h';n_1} ~ \omega_{\rm exch} (n_1,n_2),
  \\
  \lefteqn{\langle e'h'|K|eh \rangle}\nonumber\\
  &=& 2\delta_{j_s} \sum_{n_1,n_2} c^*_{e';n_1}
c_{h';n_1} c^*_{h;n_2} c_{e;n_2} ~ \omega_{\rm Coul} (n_1,n_2) \nonumber
\\
  &&+ ~ 2\delta_{j_s} \sum_{n_1,n_2} c^*_{e';n_1} c_{h';n_2}
c^*_{h;n_2} c_{e;n_1} ~\omega_{\rm exch} (n_1,n_2) \nonumber \\
  &&+ ~ 2\delta_{j_s}\sum_{n_1,n_2} c^*_{e';n_1} c_{h';n_2}
c^*_{h;n_1} c_{e;n_2} ~\omega_{\rm exch} (n_1,n_2).
  \ea

  To evaluate these integrals, we need a real-space description
of the tight-binding basis orbitals. To start, we  follow Martin {\it et al.} 
\cite{martin} and 
  model the tight-binding orbitals
with atomic Slater orbitals.\cite{slater} The Slater 
orbitals are single-exponential functions with the exponent given by the Slater
rules\cite{slater} designed to yield  
a good approximation of the effective radius and the ionization energy.
Especially, an $s^*$ orbital is  modeled as an excited $s$ orbital 
 by promoting one valence electron to the $s$ orbital of the next shell. 
The Slater orbitals are an arbitrary choice 
in the sense that they are not explicitly related to the
tight-binding parameters. However, as shown in Section~\ref{sec:results},  
even in small dots the electron-hole Coulomb interaction is \emph{not} very
sensitive to variations in the orbital integrals. 

  \emph{Onsite} Coulomb and exchange integrals, in which both orbitals are 
centered on the same atom, are calculated using a Monte Carlo
method with importance sampling for the radial integrations. The uncertainty of the Monte Carlo results is within 1\%. The angular part
is treated exactly by expansion in
 spherical harmonics.
However, the Appendix
 shows that the integral values must be considered to be uncertain
to about 20--30\% due to the arbitrariness of the orbital choice and the
effects of orthogonalization.

  \emph{Offsite exchange} integrals, where the two orbitals are centered
on two different atom sites, are negligible even for nearest-neighbor integrals. 
 These offsite exchange integrals 
decrease quickly as the distance between atom sites increases,
due to the localization and orthogonality of the orbitals. In particular, we
show in the Appendix that even nearest-neighbor exchange integrals are negligible as
an effect of orthogonalization between offsite hybrids. 

 Regarding \emph{offsite Coulomb} integrals, Leung and Whaley\cite{leung_si} estimate these integrals  
 with the Ohno formula\cite{ohno} modified to include screening: 
  \ba
  \lefteqn{\omega_{\rm Coul}(n,n') \equiv \omega_{\rm Coul} (i\gamma,i'\gamma')}
\nonumber\\
  &=& \frac{1}{\epsilon(|{\bf R}_i-{\bf R}_{i'}|,R)} \frac{1}{\sqrt{
\omega_{\rm Coul}^0 (i\gamma,i\gamma')^{-2} + |{\bf R}_i- {\bf R}_{i'}|^2 }},
\label{eq:ohno}
  \ea
  where  ${\bf R}_i$ and ${\bf R}_{i'}$ are atom site vectors. $\omega_{\rm Coul}^0 (i\gamma,i\gamma')$ is an unscreened onsite
Coulomb integral. The superscript 0 designates an unscreened quantity.
For the case of binary compounds,
 $\omega_{\rm Coul}^0(i\gamma,i\gamma')$ is replaced
 by the average over the two different atoms. 
The integrals are screened by the dielectric constant $\epsilon(|{\bf R}_i-{\bf R}_{i'}|,R)$. This screening is the only modification to the original Ohno formula. 
From here on, we will refer to this modified Ohno formula in Eq.~(\ref{eq:ohno}) simply as the Ohno formula. 

  To test the validity of the Ohno formula in the case that two orbitals
are on close atom sites, 
we calculated the offsite Coulomb integrals with a Monte Carlo
method\cite{monte} within 1\% uncertainty and compared these
values with those from the Ohno formula. 
The Ohno formula severely underestimates the offsite integrals as the distance
between two atom sites becomes small. 
For example, the Coulomb integral between the bonding orbitals (see below) from
 nearest neighbors in a Si quantum dot with radius 18.9~\AA\@
given by the Ohno formula is 0.58~eV, while the Monte Carlo calculation gives 2.35~eV. 
For next-nearest neighbors, the Ohno formula and the Monte Carlo calculation
give 0.38~eV and 0.58~eV, respectively, and for the third-nearest neighbors  0.33~eV and 0.36~eV. 

The reason that we obtain a big discrepancy between the Ohno formula and the Monte Carlo calculation for the bonding-orbital integrals is that the effective distance between the bonding-orbitals is smaller than the spacing between the nearest-neighbor atom sites. 
In fact, the spatial overlap of the bonding-orbitals is as big as that of  the orbitals on the same atom site.
In addition, the spatial dependence of the dielectric function is not fully taken into account in the Ohno formula. This approximation becomes critical when the range of variations in the dielectric function is comparable to the effective distance between orbitals. In that case, the effective dielectric function cannot be represented by $\epsilon(|{\bf R}_i-{\bf R}_j|,R)$. 
Therefore, we use the  Monte Carlo values for the onsite and the
nearest-neighbor integrals and the Ohno formula for the rest of offsite
integrals. For clarity, we summarize the methods for the computation of the Coulomb and exchange integrals in Table~\ref{tab:method}. 

   We use a size and distance dependent dielectric function to screen the Coulomb and exchange interaction of the electron-hole pair. 
The dielectric function, as a function of the separation $r$ of two 
particles  and the radius $R$ of a quantum dot,  is approximated by the
Thomas-Fermi model of Resta\cite{resta} and the Penn model generalized for quantum dots.\cite{penn}$^,$\cite{tsu} 
The separation dependence is given by the Thomas-Fermi model, while 
the size dependence is given by the Penn model.
This way of combining the two models to describe the dielectric function is taken from Ref.~\ref{ref:franceschetti}: 
  \ba
  \epsilon(r,R) &=&
  \left\{
  \begin{array}{l@{\quad,\quad}l}
  \epsilon_{\infty}^{\rm dot} (R) ~ qr_0/[\sinh q(r_0-r) + qr] & r<r_0
\\
  \epsilon_{\infty}^{\rm dot} (R) & r\geq r_0
  \end{array} \right. \nonumber \\
\label{eq:dielectric}
  \ea
where
  \ba
  \epsilon_{\infty}^{\rm dot} (R) &=& 1 + (\epsilon_{\infty}^{\rm bulk} - 1) \frac{(E_{\rm gap}^{\rm bulk} + \Delta)^2}{[E_{\rm gap}^{\rm dot} (R) + \Delta]^2}.\label{eq:infty}
  \ea
 The Thomas-Fermi wave vector $q$ is 
$(4/\pi)^{1/2}(3\pi^2n_0)^{1/6}$, where the valence electron density $n_0=32/a_0^3$ in a zincblende structure.
The screening radius $r_0$ is determined by the condition
$\sinh qr_0/qr_0 = \epsilon_{\infty}^{\rm dot} (R)$. The shift $\Delta = E_2-E_{\rm gap}^{\rm bulk}$, where $E_2$ is the energy of the first pronounced peak in the bulk absorption spectrum. The energies $E_{\rm gap}^{\rm bulk}$ and $E_{\rm gap}^{\rm dot}(R)$ are the
single-particle gaps for bulk and a dot with radius $R$, respectively. $\epsilon_{\infty}^{\rm bulk}$ is the dielectric constant for the bulk material.  

  The unscreened onsite Coulomb and exchange integrals
 for the $sp^3 s^*$ basis set are listed in
Table~\ref{tab:unscreened}, and the screened onsite Coulomb integrals 
are listed in Table~\ref{tab:screened}. The screened
nearest-neighbor Coulomb integrals are listed in
Table~\ref{tab:nnCoulomb}. 
 The
integrals with $s$ and $p$ orbitals are calculated using the four
hybridized orbitals along bonding directions, defined as:
  \ba
  |sp^3_a\rangle &=& \frac{1}{2} (|s\rangle \pm |p_x\rangle \pm |p_y\rangle
\pm|p_z\rangle), \nonumber \\
  |sp^3_b\rangle &=& \frac{1}{2} (|s\rangle \pm |p_x\rangle \mp |p_y\rangle \mp
|p_z\rangle), \nonumber \\
  |sp^3_c\rangle &=& \frac{1}{2} (|s\rangle \mp |p_x\rangle \pm |p_y\rangle \mp
|p_z\rangle), \nonumber \\
  |sp^3_d\rangle &=& \frac{1}{2} (|s\rangle \mp |p_x\rangle \mp |p_y\rangle \pm
|p_z\rangle) \label{eq:hybrid},
  \ea
 where the upper sign is for an anion and the lower sign is for a cation.
                                                                           
 As shown in Tables~\ref{tab:unscreened} and~\ref{tab:screened}, the screening
effect is significant even for onsite integrals. This is because the screening
radius $r_0$ is 2--4 \AA\@ and is similar to the effective radius
of the tight-binding basis orbitals. The comparison of the unscreened and
screened onsite integrals shows that the effective screening of these integrals
is generally about half the long range screening given by $\epsilon_{\infty}^{\rm dot}(R)$.   
Further, based on this observation,
we use half of the  long-range dielectric constant to screen the onsite
exchange integrals that we did not calculate explicitly by the Monte Carlo
method.\cite{monte_ex} 
 
\subsection{Lowest excitonic states}
\label{subsec:diagonalization}

  To obtain the excitonic states near the band edge, we diagonalize the
configuration-interaction matrix in the $|eh\rangle$ basis set given by the sum of $H_{\rm single}$ and $H_{\rm e-h}$ defined by Eqs.~(\ref{eq:single})--~(\ref{eq:h_eh}).  
We include sufficient electron and hole states in the configurations to achieve convergence of the first few excitonic states to within a few meV. The typical number of electron and hole states needed is about 10-15 each.

 There are two types of
Hamiltonians depending on the total spin of the electron-hole pair. The
Hamiltonian for a spin singlet includes both the Coulomb and
the exchange interaction, whereas the Hamiltonian
for a spin triplet has only the Coulomb interaction. By
diagonalizing these two Hamiltonians separately, we obtain a set of spin-singlet
and spin-triplet excitonic states. 
The lowest excitonic state is the lowest triplet state due to the absence of the positive
exchange interaction. 
However, since only spin-singlet states are optically allowed the   
 \emph{optical} excitonic gap
 is the energy of the lowest spin-singlet 
state.

 This order of the states can be seen most easily by applying 
  first-order perturbation theory to the electron-hole pair made  
from the highest hole state and the lowest electron state, which yields 
  \ba
  E_{\rm singlet} &=& E_e - E_h + \langle eh|J|eh \rangle + \langle
eh|K|eh \rangle. \\
  E_{\rm triplet} &=& E_e - E_h + \langle eh|J|eh \rangle.
  \ea
  The sign of $\langle eh|J|eh \rangle$ is always negative and the sign of
$\langle eh|K|eh \rangle$ is always positive. Therefore, the energy of the
lowest spin-triplet excitonic state is smaller than the energy of the
lowest spin-singlet excitonic state by $\langle eh|K|eh \rangle$ within 
first-order perturbation theory. The energy difference between
these
excitonic states is the \emph{exchange splitting}. 
Further, we denote the difference
 between the lowest spin-triplet energy and the single-particle energy gap  as 
the \emph{Coulomb shift}, which is $\langle eh|J|eh \rangle$ in first-order 
perturbation theory. The Coulomb shift is the main correction to the single-particle gap since the Coulomb interaction is roughly one order of magnitude larger than the exchange interaction. 

This simple description becomes more complicated when configuration interaction  is included due to the correlation of several electron-hole configurations near the band edges. However, the main ideas of Coulomb shift and exchange splitting are still valid. Generally, as we include more electron-hole configurations the Coulomb shift and the exchange splitting increase and converge. 

\subsection{Effective range of the exchange interaction} \label{subsec:exchange}

The long-range component of the exchange interaction was 
investigated by Franceschetti {\it et al.}\cite{franceschetti_ex} with 
the pseudo-potential method.
They show that there is a long-range component in the monopole-monopole exchange interaction for several direct gap semiconductor quantum dots.  
To verify this long-range exchange interaction 
with the tight-binding model, we follow their approach using a cutoff potential.
With the step function $\Theta(r)$, we replace the Coulomb potential with a
 cutoff potential~$\Theta(r_c-|{\bf r'}-{\bf r}|) / |{\bf r'}-{\bf r}|$ to obtain an exchange 
interaction that is a function of the cutoff distance $r_c$. 

The 
unscreened exchange interaction with the cutoff potential for electron-hole state $|eh\rangle$ is 
  \ba
  \lefteqn{\langle eh|K^0(r_c)|eh\rangle} \nonumber \\
  &=& \int\!\!\int \der^3 r' ~ \der^3 r \frac{\psi^*_{e}({\bf r'}) \psi_{h}({\bf r'})
\psi^*_{h}({\bf r}) \psi_{e}({\bf r})} {|{\bf r'}-{\bf r}|} ~ \Theta(r_c-|{\bf r'}-{\bf r}|) \nonumber \\
  &\approx& 2\!\sum_{n_1,n_2} c^*_{e;n_1} c_{h;n_1} c^*_{h;n_2}
c_{e;n_2} ~ \omega^0_{\rm Coul} (n_1,n_2) \nonumber\\
  &&\times~ \Theta(r_c-|{\bf R}_{n_1}-{\bf R}_{n_2}|) \nonumber \\
  &&+ ~ 4\!\sum_{n_1,n_2}\! c^*_{e;n_1} c_{h;n_2} c^*_{h;n_2}
c_{e;n_1} ~ \omega^0_{\rm exch} (n_1,n_2). \label{eq:cutoff}
  \ea
  where the superscript 0 refers to the unscreened interaction. 
 In line with the discrete spatial character of the tight-binding model, we make an approximation that replaces the true cutoff potential with one based on the site indices  $\Theta(r_c-|{\bf R}_{n_1}-{\bf R}_{n_2}|)$. 
If there is a long-range exchange interaction, it would stem from
the first term of Eq.~(\ref{eq:cutoff}) which includes the Coulomb integrals. 
The second term, the sum of exchange integrals, has only 
onsite integrals, since all offsite exchange integrals are negligible as shown in the Appendix.

  To understand the physical origin of the long-range exchange, we expand the exchange
  interaction $K^0$ in a 
multipole expansion:
  \ba
  \langle eh|K^0|eh \rangle &\approx& \sum_{i\not=j} \frac{q({\bf R}_i)_{eh}^*
~ q({\bf R}_j)_{eh}} {|{\bf R}_i-{\bf R}_j|} + {\cal O}(|{\bf R}_i - {\bf  
R}_j|^{-2}) \nonumber \\
  &&+ ~ \mbox{onsite interaction} \label{eq:multipole}.
  \ea
  Here the `exchange charge density' $q({\bf R}_i)_{eh}$  
at atom site ${\bf R}_i$ is a monopole moment defined as
  \ba
  q({\bf R}_i)_{eh} &\equiv& \int \der\Omega_i ~\psi_e({\bf r}) \psi_h^*({\bf r}) \nonumber \\ 
 &=&\sum_{\gamma\gamma'} c_{e;\gamma} ({\bf R}_i)
c^*_{h;\gamma'} ({\bf R}_i) \int\der^3r ~ \phi_{i\gamma}({\bf r})
\phi^*_{i\gamma'}({\bf r}) \nonumber \\
  &=& \sum_{\gamma\gamma'} c_{e;\gamma} ({\bf R}_i) c^*_{h;\gamma'} ({\bf
R}_i) \delta_{\gamma\gamma'} \nonumber \\
  &=& \sum_{\gamma}c_{e;\gamma} ({\bf R}_i) c^*_{h;\gamma} ({\bf R}_i),
\label{eq:charge}
  \ea
 where $\int \der \Omega_i$ is defined to integrate only the orbitals on the atom site ${\bf R}_i$. 
For clarity, the tight-binding orbital index $n$ is specifically replaced by the atom-site index $i$ (or ${\bf R}_i$) and
 the orbital-type index $\gamma$.  
  Note that the final expression for $q({\bf R}_i)_{eh}$ has a sum over only
one orbital-type index due to the assumed orthogonality of the tight-binding basis
orbitals. The distribution of the exchange charge density  determines the long-range character of the exchange interaction. 
If the exchange charge density is zero, that is, the electron and hole states are locally (onsite) orthogonal, there is no exchange interaction beyond the onsite contribution according to Eq.~(\ref{eq:multipole}). In contrast, if the exchange charge density is nonzero due to the local nonorthogonality of the electron and hole wave functions from site to site, 
a long-range exchange interaction is caused by the monopole-monopole interaction. 

\section{Results} \label{sec:results}

\subsection{Real-space description of basis orbitals} \label{subsec:basis}
  The empirical tight-binding model has an inherent difficulty concerning
the tight-binding basis orbitals. 
The real-space description of the basis
orbitals is not provided since the 
tight-binding matrix elements are determined by fitting to the bulk band
structure. 
However, to include electron-hole correlations the electron-hole Coulomb and exchange matrix elements need to be calculated, which requires an explicit choice of 
real-space basis orbitals.
Since this choice is largely arbitrary in the sense that there is no way to connect the chosen basis orbitals to the empirically chosen tight-binding parameters, 
we need to test to what degree the
choice of orbitals affect the electron-hole Coulomb interaction.

  We perform this test by scaling the onsite Coulomb and exchange integrals
 from the values listed in Tables~\ref{tab:unscreened} and~\ref{tab:screened}\@. 
This scaling scheme is an indirect way of 
testing the sensitivity on the real-space description. 
 New offsite Coulomb
integrals are calculated by replacing the unscreened onsite integrals with
the scaled ones in the Ohno formula,
Eq.~(\ref{eq:ohno}).
Note that the offsite Coulomb integrals are not directly scaled by the
same factor as the onsite integrals, but change only indirectly 
through the scaled unscreened onsite integrals in the Ohno formula.
 Therefore, this scheme is in effect changing atomistic details of the basis orbitals.
 
  Figure~\ref{fig:sen} shows the variation of the Coulomb interaction between
the highest hole state and the lowest electron state with the scaling of 
the onsite
Coulomb and exchange integrals. It shows
that as the dot size increases, the sensitivity of the Coulomb interaction
to the onsite integrals decreases. 
For example, if the onsite integrals are reduced by 50\%, the reduction in the Coulomb energy varies from 20\% in the smallest shown dot to only 5\% at 30~\AA\@ radius. 
Since the contribution from the
onsite integrals decreases as the dot size increases, the
specific model of the real-space functions for the basis orbitals is less
critical for larger dots.

  We can explain this effect by a closer look at the Ohno formula
for the offsite integrals.  In the limit of large
distances between two atom sites, the offsite integrals
become independent of the onsite integral values.
The unscreened offsite integral in the limit of large
distance between the two atom sites is
  \ba
  \lefteqn{\omega^0_{\rm Coul} (i\gamma,i'\gamma')}\nonumber \\
  &\approx& \frac{1} {|{\bf R}_i- {\bf R}_{i'}|}- \frac{1} 
  {2\omega^0_{\rm Coul} (i\gamma,i\gamma')^2 ~|{\bf R}_i-{\bf R}_{i'}|^3}.
  \label{eqn:ohno_long}
  \ea
  Therefore, the offsite integrals become a 
point-charge interaction, making the atomic-scale details of the basis orbitals 
irrelevant in this limit. 

The significance of Figure~\ref{fig:sen} is that it quantifies this qualitative explanation for decreasing sensitivity with increasing dot sizes. For example, with a targeted 10\% accuracy in the Coulomb interaction, only a 50\% accuracy for the onsite integrals is needed for dots of radius larger than 30~\AA, while for dots of 10~\AA\@ radius only 20\% error can be afforded in the onsite integrals.
The discussion in the Appendix shows that the dominant integrals must be
considered uncertain to about 20-30\%. The tight-binding description of
correlation effects can therefore be considered reliable for dots with radii
larger than 15--20 \AA.
 
\subsection{Excitonic states near the band edge} \label{subsec:exciton}
  We apply the tight-binding configuration interaction scheme 
described in Section~\ref{subsec:hamiltonian} to Si, InAs, and CdSe quantum dots
in order to calculate the lowest
excitonic states near the band edge.
We have included sufficient electron and hole states to converge the energies
to within a few~meV.

  For Si, we first used the tight-binding parameters of 
Vogl {\it et al.}\cite{vogl} 
to determine
the single-particle Hamiltonian matrix elements. 
As shown in Figure~\ref{fig:si_gap}, 
the excitonic gap using their tight-binding parameters gives a
discrepancy as large as 0.3~eV compared with experimental data.\cite{wolkin}
The tight-binding parameters of Vogl {\it et al.} necessarily give 
poor effective masses
since the parameters are determined by fitting  to
the energies of only  the $\Gamma$ and $X$ points in the bulk band structure. 
The resulting effective masses are listed in Table~\ref{tab:effectivemass}.
The comparison with the experimental values listed in Table~\ref{tab:effectivemass} shows that their parameters fail to produce good effective masses especially for the transverse effective masses at the lowest conduction energy near $X$.

  To improve the effective masses,  
 we replace the parameter set of Vogl {\it et al.}
 with our parameter
set listed in Table~\ref{tab:si_tb}. 
Our parameter set is optimized with a genetic algorithm by fitting 
the effective masses as well as the energies at high 
symmetry points of the bulk band structure.\cite{klimeck} 
The resulting effective masses are listed in Table~\ref{tab:effectivemass}.
One important note is that we
use two different parameter sets to separately optimize the
electron and hole single-particle states. 
Good effective masses are impossible to obtain simultaneously for both the conduction and the valence band of Si with one set of parameters within the nearest-neighbor $sp^3s^*$ tight-binding model (see Ref.~\ref{ref:klimeck}).
Consequently, the electron and hole single-particle wave functions, being  generated
from different Hamiltonians, are not orthogonal.
However, even though the
orthogonality has not been enforced, the overlaps between the different
electron and hole wave functions are at most 0.001.  
Thus, we can use these two different
parameter sets to verify how important a role the effective masses play
in the electronic properties of the quantum dots. 

  Figure~\ref{fig:si_gap}  shows the improved excitonic gaps
with our parameter set. To further examine the effect of changing  
parameters, we can compare the electron-hole interaction energies 
 with our parameters to those energies obtained with the parameters of Vogl {\it et al.}  
In particular, we compare the Coulomb shift, the energy difference
between the single-particle gap and the lowest triplet excitonic energy. 
As shown in the inset in 
Figure~\ref{fig:si_gap},
the Coulomb shifts from the two parameter sets
are very similar.
This insensitivity indicates that the better description of the
excitonic gap with our parameter set is mainly due to  the better
single-particle eigenvalues and not from a change in the Coulomb matrix
elements. 

  To study direct gap semiconductors, we choose InAs and
CdSe spherical quantum dots.
The InAs tight-binding parameters are generated using the genetic algorithm
approach,\cite{klimeck} fitting band
gaps and effective masses at $\Gamma$ as well as possible, but neglecting
spin-orbit coupling. These parameters are listed in Table~\ref{tab:inas_tb}.
The resulting effective masses with these parameters are $m_c=0.024$, $m_{vl}=0.025$, and $m_{vh}=0.405$, where $m_c$ is the effective mass of the lowest conduction band at $\Gamma$, and $m_{vl}$ and $m_{vh}$ are defined as in Table~\ref{tab:effectivemass}. 
The tight-binding parameters for CdSe are taken from Ref.~\ref{ref:lippens}.

Figures~\ref{fig:cdse_gap} and~\ref{fig:inas_gap} show the resulting 
excitonic gaps versus the dot radius. 
  For CdSe quantum dots, our excitonic gaps are in good
agreement with optical gaps measured by photoluminescence excitation\cite{norris} (PLE). 
We also plot the energy gaps measured by scanning tunneling spectroscopy (STM)
on a single quantum dot.\cite{alperson} 
The STM gaps are obtained from the difference between the first prominent peaks of the
tunneling $dI/dV$ spectra with positive and negative bias voltages, respectively.
Since the STM experiment applies bias voltages to add or subtract electrons from the quantum dots, this experiment measures quasiparticle energies. For a
finite system, these quasiparticle energies include the (positive)
polarization energy between the particle and the image charges on the
surface.
The polarization energy is roughly $2(1/\epsilon_{\rm out}-1/\epsilon_{\rm in})/R$. The small difference between our single-particle gaps and the STM gaps
therefore suggests that the dielectric constant $\epsilon_{\rm out}$ of the surrounding material is
relatively close to the dielectric constant $\epsilon_{\rm in}$ in the dots.
The results of pseudo-potential calculations\cite{franceschetti} are also
plotted in the figures for comparison. 

For InAs quantum dots, 
there is a significant discrepancy as large as 0.2~eV between
our $sp^3s^*$ tight-binding excitonic gaps and PLE gaps.\cite{banin}
Eight-band effective-mass calculations \cite{banin} and pseudo-potential calculations\cite{williamson} also fail to
describe the experimental data, especially the lack of significant
curvature. The recent results of Allan {\it et al.} show that the inclusion of $d$
orbitals and spin-orbit interaction raises the $sp^3s^*$ results by almost the
needed 0.2 eV. However, their results do not include the Coulomb shift and
should therefore be shifted down by 200--50 meV as the dot size increases.
Recent STM measurements are also plotted in Fig.~\ref{fig:inas_gap}. 
It is consistent with the larger dielectric constant of
InAs that the STM results in this case is well above the other curves by an
amount similar to the Coulomb shift.

\subsection{Effective range of the exchange interaction}
\label{subsec:exchange2}

  One of the interesting issues related to the exchange interaction of the
electron-hole pair is its effective range. 
Motivated by the work of Franceschetti {\it et al.},\cite{franceschetti_ex} 
we calculate the unscreened
exchange interaction defined in Eq.~(\ref{eq:cutoff}) as a function of the
cutoff distance $r_c$ to determine the effective range of the exchange interaction.
As the cutoff distance increases for a given electron-hole configuration, the exchange interaction eventually saturates to a final value. If this saturation occurs over just a few atomic sites, we call it short-ranged, while long-ranged exchange implies that the saturation occurs over distances comparable to the dot size.   
For Si, Figure~\ref{fig:si_exch} shows that for the configuration with the highest
hole state and the lowest electron state 
there is a region of strong oscillations below a cutoff distance of
15~\AA. 
The strong oscillations
are due to the phase difference between
the electron and hole states stemming from their different locations in $k$-space for an
indirect-gap material. 
The oscillations die out beyond a cutoff
distance of about 15~\AA\@, suggesting that the
effective range of the exchange interaction in Si quantum dots is around
15~\AA\@ regardless of dot size. This short-ranged and oscillatory behavior is universal within the configurations near the band edges.

For the direct-gap InAs and CdSe 
quantum dots,  we calculate the unscreened exchange interaction for several
of the lowest electron-hole configurations. 
We label the electron and hole states by the dominant
angular momentum character of their `envelope functions.' Here, the envelope function
is defined to be the coefficient of the dominant basis orbital. 
The $s$
and $p$ basis orbitals are typically dominant in the electron and hole states,
respectively.  In our calculations, a 
$p$-like hole\cite{notation_eh} is the highest
hole state and an $s$-like hole is the second-highest hole state.
This order is opposite to that of 
pseudo-potential theory.\cite{franceschetti} 
However, it is possible that the spin-orbit coupling, which is not included 
in this work, can affect the order of these hole states.  
 
  As shown in Figures~\ref{fig:cdse_exch} and~\ref{fig:inas_exch},
direct-gap quantum dots show a qualitatively different behavior
of the exchange interaction with respect to the cutoff distance from 
the behavior for Si. First, since there is no overall phase
difference between the electron and hole states, there
is no region with oscillations for small cutoff distances. Second,
the exchange interaction for a particular electron-hole pair can grow 
continuously up to the dot radius.
The figures show that the exchange interaction of
direct-gap materials is generally 
long ranged, extending over the whole dot.

To understand why some electron-hole configurations have a slowly varying  
long-range exchange interaction, we analyze 
the long-range component
by a multipole-expansion as
written in Eq.~(\ref{eq:multipole}).  The leading term of the 
long-range exchange interaction
is the monopole-monopole interaction.
Therefore, the distribution of the monopole moment, or the `exchange
charge density' defined in Eq.~(\ref{eq:charge}), determines the
range of the long-range exchange interaction.

The exchange charge density of an electron-hole pair has zero
total charge due to the orthogonality between the electron and hole
wave functions. There are two ways to satisfy this condition: the electron and
hole states are either locally orthogonal, which is enforced in effective
mass theory due to the orthogonality between the Bloch functions of the valence and conduction bands, or globally orthogonal which is possible in the atomistic
pseudo-potential and tight-binding theories. 
If the former is true, the exchange charge density would be zero
at each site
and there would be no  monopole--monopole interaction. 
That would make the exchange interaction of the electron-hole pair short ranged.  
By contrast, without onsite orthogonality the exchange charge density has
nonzero values, causing monopole-monopole interactions that lead to
significant long-range exchange interactions. 

To show that the character of the orthogonality of the electron-hole configuration
determines the long-range behavior of the exchange interaction, we plot
in Figure~\ref{fig:cdse_charge}
the exchange charge density of two electron-hole configurations in CdSe that have
a long-range exchange interaction in Figure~\ref{fig:cdse_exch}. 
Figure~\ref{fig:cdse_charge} shows the exchange charge density of (a) the
$s$-like electron and $s$-like hole configuration, and of (b) the $p$-like electron
and $s$-like hole configuration in a plane going through the center of the dot for CdSe
with radius 20.1~\AA\@. This figure shows that there is no local
orthogonality between the electron and hole wave functions.
The orthogonality of the electron and hole wave functions are instead satisfied
by a $p$-like global oscillation (case a) or a 2$s$-like global oscillation
(case b). 
These shapes of the global oscillations explain why the exchange
interaction  has growing and  decaying regions over global distances
 as shown in Figures~\ref{fig:cdse_exch} and~ \ref{fig:inas_exch}.           
Those electron-hole configurations that do not have a long-range exchange
interaction have a much smaller exchange charge density than those configurations
that do have the long-range exchange interaction. These results show that
local nonorthogonality of the electron and hole wave functions leads to a strong
monopole--monopole interaction, and that the global variations in the
exchange interaction depend on the particular way in which the exchange
charge density globally sums to zero for a specific electron-hole configuration.

\section{Summary} \label{sec:summary}

  We use tight-binding wave functions to calculate electron-hole states
near the band edge for both direct-gap and indirect-gap
quantum dots. First, we examined to what
degree the model of the real-space atomic basis orbitals affects the
electron-hole Coulomb interaction. We find that the sensitivity of
the Coulomb interaction to the real-space description of the basis orbitals
decreases quickly as the dot size increases. Our results shows that 
tight-binding descriptions of electron-hole Coulomb interactions in
quantum dots should be reliable for dots larger than about 
15--20~\AA\@ radius  even for simple models for the basis orbitals. 
More detailed calculations of basis orbitals are required for smaller dots.
 
For excitonic gaps, we obtained good agreement with    
recent experiments for both Si and CdSe quantum dots. 
However, 
the gaps for InAs quantum dots agree less well with experiment.
Especially for Si, we improved the agreement with experimental data for the
excitonic gap  by optimizing the tight-binding parameters
to give better effective masses 
compared to the parameters of Vogl {\it et al.}
We also showed that, in contrast to the electron and hole single-particle 
energies, the electron-hole Coulomb interaction 
is not very sensitive to the choice of parameters. 

  Finally, we studied the effective range of the exchange
interaction. Replacing the Coulomb
potential with a cutoff potential, we demonstrated the
dependence of the exchange interaction on the cutoff radius. 
The existence of a global exchange charge density 
of an electron-hole pair, due to the lack of onsite orthogonality,
causes the exchange interaction to be long ranged in
direct-gap materials. For indirect materials, the calculations for Si show
that the exchange interaction is oscillatory and has a range of about 15~\AA.

\section*{Acknowledgments} \label{sec:acknw} 
We thank Jeongnim Kim for many helpful discussions. This work was supported
by NSF (PHY-9722127) and by the NCSA and OSC supercomputer centers.
Part of the work described in this paper was carried out by the Jet Propulsion Laboratory, 
California Institute of Technology under a contract with the National Aeronautics and Space
Administration. The supercomputer used in this investigation at JPL was 
provided by funding from the NASA Offices of Earth Science, Aeronautics, and Space Science.

\appendix
\section*{} \label{sec:appendix}
Since we use empirical tight-binding wave functions, the choice of specific
atomistic orbitals for matrix-element calculations is largely arbitrary.
The results presented in the main text were based on Coulomb and exchange
integrals calculated with Slater's atomic orbitals,
obtained from Slater's rules.\cite{slater} In addition, we neglected all offsite exchange
integrals. An alternative, for unscreened integrals, is to use one of the
standard Quantum Chemistry Gaussian-based commercial packages. However,
screened matrix elements cannot be obtained in this way, since there is no
way in these codes to include a spatially varying screening function.

Although we cannot obtain screened integrals, two important questions can be
answered by a comparison between integrals from Gaussian-type orbitals (GTO)
and Slater's orbitals (SO): what is the typical variation in the integral
values for two reasonable choices of orbitals; and what is the effect of
using nonorthogonal bond hybrids rather than properly orthogonalized hybrids?
The underlying assumption in the tight-binding approach is that the orbitals
on different sites are orthogonal. 

Table~\ref{tab:A1} shows a comparison between orthogonal GTO (O-GTO),
nonorthogonal GTO (NO-GTO), and 
nonorthogonal SO (NO-SO) for onsite Coulomb and exchange integrals.
Typically, the NO-SO and NO-GTO Coulomb integrals differ by 10\%, whereas the
(an order of magnitude smaller) exchange integrals differ by 20--50\%.
Orthogonalization generally gives an additional 10\% change. The use of
nonorthogonal Slater orbitals can therefore be estimated to imply 20\%
overall uncertainty in the onsite integrals.

A similar comparison for nearest-neighbor integrals is shown in Table~\ref{tab:A2}.
Here the difference between NO-GTO and NO-SO is less than 10\%, but
orthogonalization can yield a lowering of up to 30\% in the Coulomb integrals
between bonding orbitals. The most dramatic effect, however, is that the
exchange integrals essentially become negligible when orthogonalized hybrids
are used. Notably, nonorthogonal hybrids cannot be used for the bonding-bonding
offsite exchange integrals, since these integrals are quite large without
orthogonalization but are reduced by a factor of 20--30 after orthogonalization.

In conclusion, the dominant Coulomb integrals obtained from the Slater
orbitals can be considered accurate only to
20--30\% due to the sensitivity to different functional representations and to
effects of orthogonalization. Further, proper orthogonalization reveals that
all offsite exchange integrals can be neglected, including those between
bonding hybrids.

\newpage
\begin{table}
  \caption{Methods for the computations of the Coulomb and exchange integrals with respect to a site-to-site distance. NN stands for nearest neighbors. }
  \begin{tabular}{cccc}
 & onsite & NN & beyond NN \\
  \hline
  Coulomb & Monte Carlo & Monte Carlo & Ohno formula \\
  exchange & Monte Carlo & neglected & neglected 
  \label{tab:method}
  \end{tabular}
\end{table}
 
\begin{table}
  \caption{Onsite \emph{unscreened} Coulomb integrals, Eq.~(\ref{eq:Coul}), and exchange integrals, Eq.~(\ref{eq:exch}), of 
the $sp^3s^*$ basis set, in units of eV. Integrals for the
$sp^3$ orbitals are calculated based on the hybridized orbitals along the
bonding directions, Eq.~(\ref{eq:hybrid}).} 
  \begin{tabular}{clcccc}
  Element & Integral & $ (sp^3_a,sp^3_a) $ & $ (sp^3_a,sp^3_b) $ & $ (sp^3_a,s^*) $ & $
(s^*,s^*)$ \\
  \hline
  Si & Coulomb & 11.88 & 8.49 & 2.87 & 2.27 \\
  Si & exchange & 11.88 & 0.78 & 0.024 & 2.27 \\
  In & Coulomb & 7.82 & 5.67 & 2.30 & 1.36 \\
  In & exchange & 7.82 & 0.47 & 0.024 & 1.36 \\
  As & Coulomb & 12.13 & 9.26 & 2.34 & 1.73 \\
  As & exchange & 12.13 & 0.47 & 0.028 & 1.73 \\
  Cd & Coulomb & 6.59 & 5.06 & 1.77 & 1.44 \\
  Cd & exchange & 6.59 & 0.74 & 0.017 & 1.44 \\
  Se & Coulomb & 12.85 & 9.70 & 2.97 & 2.38 \\
  Se & exchange & 12.85 & 0.95 & 0.030 & 2.38
  \label{tab:unscreened}
  \end{tabular}
\end{table}

\begin{table}
  \caption{Onsite \emph{screened} Coulomb integrals, Eq.~(\ref{eq:Coul}), of the
$sp^3$ hybridized orbitals and $s^*$ orbital for the Si dot with radius R=18.9~\AA, InAs with R=21.1~\AA, and CdSe with R=21.1~\AA, in units of eV. The integrals are screened by the dielectric function in Eq.~(\ref{eq:dielectric}), which is a function of electron-hole separation and of the radius of the quantum dot. For
comparison, the values in parentheses are the integrals obtained from full
screening with the dielectric constant in the long-distance limit
$\epsilon^{\rm dot}_{\infty}(R)$.} 
  \begin{tabular}{ccccc}
  Element & $(sp^3_a,sp^3_a) $ & $ (sp^3_a,sp^3_b) $ & $ (sp^3_a,s^*) $ & $
(s^*,s^*)$ \\
  \hline
  Si & 3.16 (1.26) & 1.67 (0.91) & 0.32 (0.31) & 0.26 (0.24) \\ 
  In & 1.26 (0.80) & 0.77 (0.58) & 0.24 (0.24) & 0.17 (0.14) \\
  As & 2.33 (1.24) & 1.33 (0.95) & 0.28 (0.24) & 0.23 (0.18) \\ 
  Cd & 1.86 (1.29) & 1.20 (0.99) & 0.35 (0.35) & 0.28 (0.28)\\
  Se & 4.74 (2.51) & 2.76 (1.89) & 0.58 (0.58) & 0.48 (0.46) 
  \label{tab:screened}
  \end{tabular}
\end{table}

\begin{table}
  \caption{Nearest-neighbor \emph{screened} Coulomb integrals, Eq.~(\ref{eq:Coul}), of 
 the $sp^3$ hybridized orbitals for the Si dot with radius R=18.9~\AA, InAs with R=21.1~\AA, and CdSe with R=21.1~\AA, in units of eV. The integrals are screened by the dielectric function in Eq.~(\ref{eq:dielectric}). The values given by the Ohno formula Eq.~(\ref{eq:ohno}) are listed within parentheses.} 
  \begin{tabular}{lcccc}
  & Si bonding & Si non-bonding \\
  \hline
  Si bonding & 2.35 (0.58) & 0.95 (0.53) \\
  Si non-bonding & 0.95 (0.53) & 0.55 (0.53) \\
  \hline
  & In bonding & In non-bonding \\
  \hline
  As bonding & 1.43 (0.54) & 0.61 (0.54) \\
  As non-bonding & 0.81 (0.53) & 0.42 (0.53)\\ 
  \hline
  & Cd bonding & Cd non-bonding \\
  \hline
  Se bonding & 2.41 (1.03) & 1.03 (1.02) \\ 
  Se non-bonding & 1.62 (1.02) & 0.79 (1.00)  
  \label{tab:nnCoulomb}
  \end{tabular}
\end{table} 

\begin{table}
   \caption{Effective masses of Si with the tight-binding parameters of Vogl {\it et al.} and our parameters in Table~\ref{tab:si_tb}, in units of the free electron mass. $m_{cl}$ and  $m_{ct}$ denote the longitudinal and transverse effective masses at the lowest conduction energy near $X$. $m_{vl}$ and $m_{vh}$ are the effective masses at $\Gamma$ of the two highest valence bands with a light mass and a heavy mass, respectively. 
The hole masses are averages of the three directions given in Ref.~\ref{ref:klimeck}.
The cyclotron resonance data are taken from Ref.~\ref{ref:springer}.}
  \begin{tabular}{lcccc}
  & $m_{cl}$ & $m_{ct}$ & $m_{vl}$ & $m_{vh}$ \\
  \hline
  Vogl {\it et al.} & 0.73 & 1.61 & 0.18 & 0.39 \\
  Our parameters & 0.91 & 0.30 & 0.15 & 0.55 \\
  Cyclotron resonance & 0.92 & 0.19 & 0.15 & 0.54  
  \label{tab:effectivemass}
  \end{tabular}
\end{table}

\begin{table}
	\caption{Tight-binding parameters for electron and hole
states of Si, in units of eV.  The notation\cite{notation} of 
Vogl {\it et al.}\cite{vogl} is used. }
	\begin{tabular}{ccccc}
	& $E(s)$ & $E(p)$ & $E(s^*)$ & $V(s,s)$\\
	\hline
      electron & -3.060 & 1.675 & 4.756 & -8.114 \\ 
      hole & -4.777 & 1.674 & 8.697 & -8.465 \\ 
      Vogl {\it et al.} & -4.200 & 1.715 & 6.685 & -8.300 \\
        \hline
         & $V(x,x)$ & $V(x,y)$ & $V(s,p)$ & $V(s^*,p)$ \\
	\hline
       electron &  1.675 & 21.838 & 8.236 & 5.994 \\
	hole & 1.674 & 4.919 & 5.724 & 6.133 \\
        Vogl {\it et al.} &  1.715 & 4.575 & 5.729 & 5.375  
	\label{tab:si_tb}
	\end{tabular}
\end{table}

\begin{table}
	\caption{Tight-binding parameters for InAs, in units of eV. 
         The notation\cite{notation} of Vogl {\it et al.}\cite{vogl} is used.         
         Indices $a$ and $c$ refer to anion and cation, respectively.}
	\begin{tabular}{ccccc}
	$E(s,a)$ & $E(p,a)$ & $E(s,c)$ & $E(p,c)$ & \\
	\hline
         -8.419 & 0.096 & -2.244 & 0.096 & \\ 
        \hline 
       $E(s^*,a)$ & $E(s^*,c)$ & $V(s,s)$ & $V(x,x)$ & $V(x,y)$ \\
	\hline
        12.147 & 7.485 & -4.267 & 1.427 & 5.356 \\
        \hline
        $V(sa,pc)$ & $V(sc,pa)$ & $V(s^*a,pc)$ & $V(s^*c,pa)$ \\ 
        \hline
	4.409 & 5.326 & 5.846 & 4.594 & 
	\label{tab:inas_tb}
	\end{tabular}
\end{table}

\begin{table}
  \caption{Onsite unscreened Coulomb and exchange integrals with: (O-GTO) L\"owdin orthogonalized
  Gaussian-type hybrids; (NO-GTO) nonorthogonal Gaussian-type hybrids; and (NO-SO)
nonorthogonal Slater orbitals. The GTO integrals were calculated with the MOLPRO\cite{molpro} package 
using the atomic pseudo-potentials from the Los Alamos group.\cite{lanl}
The SO integrals are from our Monte Carlo calculations. The hybrids $a$ and $b$ are the
  ones defined as $sp^3_a$ and $sp^3_b$ in Eq.~(\ref{eq:hybrid}).
}
  \begin{tabular}{lrrr}
         & O-GTO     &  NO-GTO    &   NO-SO \\
  \hline
$\omega^0_{\rm Coul}(a,a)$ of Si  &11.95 & 11.65 & 11.91\\
$\omega^0_{\rm Coul}(a,b)$ of Si  & 9.44 &  8.85 &  9.00\\
$\omega^0_{\rm exch}(a,b)$ of Si  & 1.06 &  0.91 &  0.73\\
$\omega^0_{\rm Coul}(a,a)$ of In  & 7.90  & 8.52 &  7.82\\
$\omega^0_{\rm Coul}(a,b)$ of In  & 6.73 &  6.54 &  5.67\\
$\omega^0_{\rm exch}(a,b)$ of In  & 0.77 &  0.67 &  0.47\\
$\omega^0_{\rm Coul}(a,a)$ of As  &12.99 & 12.57 & 12.13\\
$\omega^0_{\rm Coul}(a,b)$ of As  &10.00 &  9.54 &  9.26\\
$\omega^0_{\rm exch}(a,b)$ of As  & 1.08 &  0.99 &  0.47\\
$\omega^0_{\rm Coul}(a,a)$ of Cd  & 7.09 &  7.81 &  6.59\\
$\omega^0_{\rm Coul}(a,b)$ of Cd  & 6.08 &  5.98 &  5.06\\
$\omega^0_{\rm exch}(a,b)$ of Cd  & 0.70 &  0.61 &  0.74\\
$\omega^0_{\rm Coul}(a,a)$ of Se  &14.14 & 13.73 & 12.85\\
$\omega^0_{\rm Coul}(a,b)$ of Se  &10.80 & 10.39 &  9.70\\
$\omega^0_{\rm exch}(a,b)$ of Se  & 1.15 &  1.08 &  0.90
\label{tab:A1}
\end{tabular}
\end{table}

\begin{table}
  \caption{Nearest-neighbor unscreened Coulomb and exchange integrals with: (O-GTO) L\"owdin
orthogonalized
  Gaussian-type hybrids; (NO-GTO) nonorthogonal Gaussian-type hybrids; and (NO-SO)
nonorthogonal Slater orbitals. The GTO integrals were calculated with the MOLPRO\cite{molpro}
package using the pseudo-potentials from the Los Alamos group\cite{lanl}
for a two-atom molecule with a bond
length given by the bulk value.
The SO integrals are from our Monte Carlo calculations. The indices $B$ and $N$ designate 
the bonding and non-bonding hybrids, respectively.
}

  \begin{tabular}{lrrr}
Si          & O-GTO     &  NO-GTO    &   NO-SO \\
  \hline
$\omega^0_{\rm Coul}(B,B)$ &  8.04 & 10.01 & 10.60\\
$\omega^0_{\rm Coul}(B,N)$ &  5.96 &  6.65 &  6.78\\
$\omega^0_{\rm Coul}(N,N)$ &  4.64 &  4.67 &  4.89\\
$\omega^0_{\rm exch}(B,B)$ &  0.27 &  6.20\\
$\omega^0_{\rm exch}(B,N)$ &  0.11 &  0.43\\
$\omega^0_{\rm exch}(N,N)$ &  0.04 &  0.32\\
  \hline
InAs          & O-GTO     &  NO-GTO    &   NO-SO \\
  \hline
$\omega^0_{\rm Coul}(B,B)$ &  6.94 &  8.77 &  9.06\\
$\omega^0_{\rm Coul}(B,N)$ &  5.50 &  6.39 &  6.59\\
$\omega^0_{\rm Coul}(N,B)$ &  5.02 &  5.42 &  5.43\\
$\omega^0_{\rm Coul}(N,N)$ &  4.12 &  4.18 &  4.24\\
$\omega^0_{\rm exch}(B,B)$ &  0.28 &  4.90\\
$\omega^0_{\rm exch}(B,N)$ &  0.16 &  0.59\\
$\omega^0_{\rm exch}(N,B)$ &  0.04 &  0.18\\
$\omega^0_{\rm exch}(N,N)$ &  0.04 &  0.29\\
  \hline
CdSe          & O-GTO     &  NO-GTO    &   NO-SO \\
  \hline
$\omega^0_{\rm Coul}(B,B)$ &  6.94 &  8.77 &  9.06\\
$\omega^0_{\rm Coul}(B,B)$ &  6.89 &  8.66 &  8.74\\
$\omega^0_{\rm Coul}(B,N)$ &  5.66 &  6.62 &  6.84\\
$\omega^0_{\rm Coul}(N,B)$ &  4.85 &  5.16 &  5.01\\
$\omega^0_{\rm Coul}(N,N)$ &  4.06 &  4.11 &  4.13\\
$\omega^0_{\rm exch}(B,B)$ &  0.27 &  4.35\\
$\omega^0_{\rm exch}(B,N)$ &  0.19 &  0.69\\
$\omega^0_{\rm exch}(N,B)$ &  0.03 &  0.13\\
$\omega^0_{\rm exch}(N,N)$ &  0.04 &  0.24 &
  \label{tab:A2}
  \end{tabular}
\end{table}

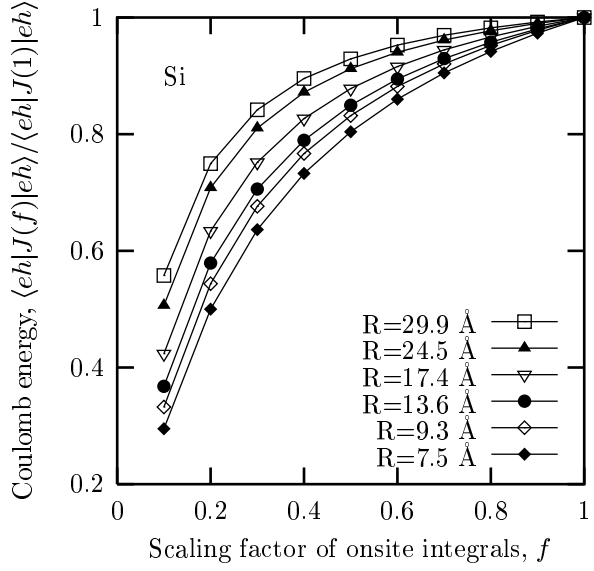
\begin{figure}
  \caption{Coulomb energy $\langle eh|J(f)|eh \rangle$ as a function of scaling factor $f$ of the 
onsite integrals. The Coulomb energy with the highest hole wave function 
and the lowest electron wave function $|eh\rangle$ is shown for various radii of Si
spherical quantum dots, with the onsite integrals scaled by the factor $f$. That is 
$\omega\rightarrow f\omega$
from the values in Tables~\ref{tab:unscreened} and \ref{tab:screened}.
The offsite integrals are only indirectly scaled through the
onsite integrals in the Ohno formula, Eq.~(\ref{eq:ohno}).
The Coulomb energy is normalized by its value at $f = 1$. 
As the dot size
increases, the Coulomb energy becomes significantly less sensitive to
variations in the onsite integrals.}
  \label{fig:sen}
\end{figure} 

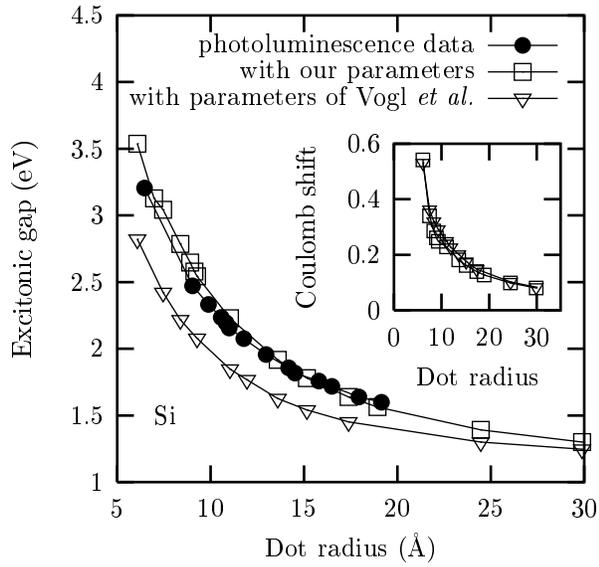
\begin{figure}
  \caption{Excitonic gap of Si spherical quantum dots as a function of the 
dot radius. The photoluminescence data are taken from Ref.~\ref{ref:wolkin}.
The other two sets of excitonic gaps are calculated with 
the tight-binding parameters of Vogl {\it et al.}\cite{vogl} and 
our parameters in Table~\ref{tab:si_tb}, respectively. 
Our parameters give significantly better agreement with experiment 
than the parameters of Vogl {\it et al.} 
This good agreement is due to the improved effective 
masses obtained with our optimized parameters.
Inset: Coulomb shift versus the dot radius. 
The Coulomb shift does not vary much between the parameter sets.}
  \label{fig:si_gap}
\end{figure}  

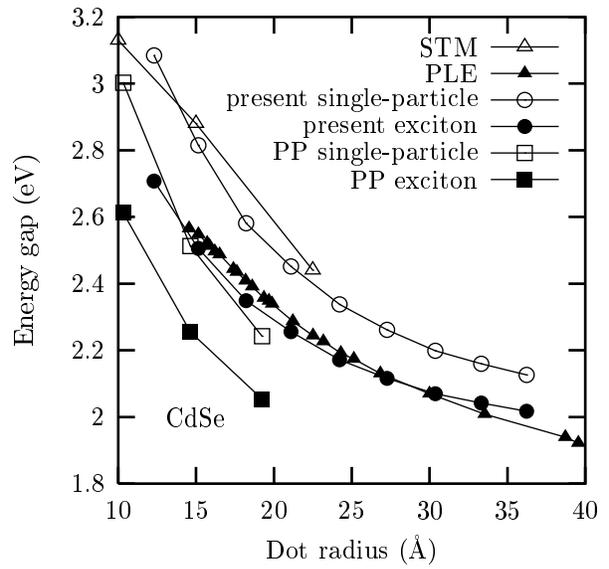
\begin{figure}
  \caption{Excitonic gap and single-particle gap of CdSe spherical quantum dots as a
function of the 
dot radius. The photoluminescence excitation (PLE) gaps are  
 taken 
from Ref.~\ref{ref:norris}. The scanning tunneling spectroscopy (STM) gaps are obtained  
from recent STM tunneling $dI/dV$ spectra.\cite{alperson}$^,$\cite{stm_data} 
 The excitonic gaps of the pseudo-potential (PP) calculations\cite{franceschetti} are
about 0.15~eV lower than 
the PLE gaps.
 Our excitonic gaps are in good agreement with the PLE gaps. 
The small difference between our single-particle gaps and the STM quasiparticle gaps
indicates that the quasiparticle polarization energy is small for these dots.
}
  \label{fig:cdse_gap}
\end{figure}     

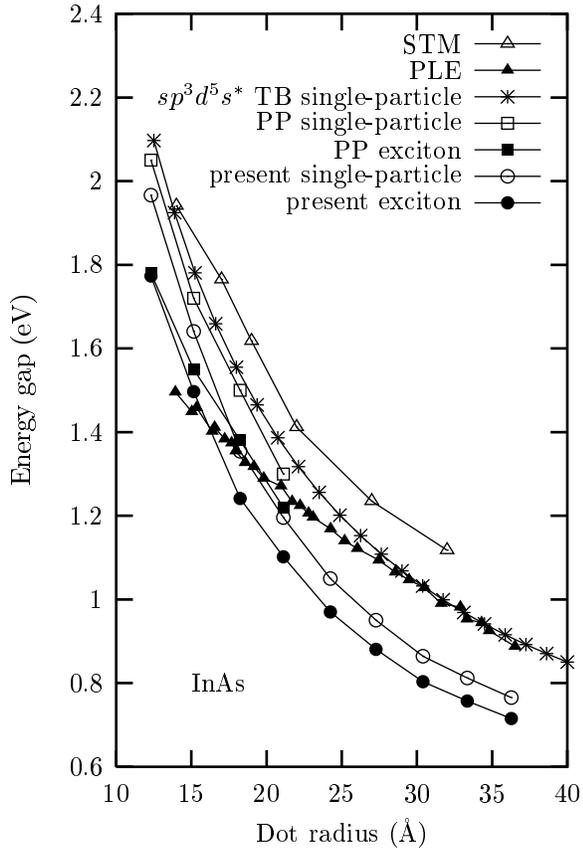
\begin{figure}
  \caption{Excitonic gap and single-particle gap of InAs spherical quantum
dots as a function of the dot radius. The measured PLE gaps are taken from 
Ref.~\ref{ref:banin}.
The STM gaps are obtained from the tunneling spectra of Millo {\it et
al.}\cite{millo}$^{,}$\cite{stm_data} 
The pseudo-potential gaps (PP) are from Ref.~\ref{ref:williamson}.
The $sp^3d^5s^*$ tight-binding (TB) single-particle gaps are plotted using
the fitting parameters of Allan {\it et al.}\cite{allan}
The inclusion of $d$ orbitals and spin-orbit coupling raises the
gaps as much as 0.2~eV in comparison with our $sp^3s^*$ model.
It is not understood why the experimental curve is so much flatter than
the theoretical curves.}
  \label{fig:inas_gap}
\end{figure} 

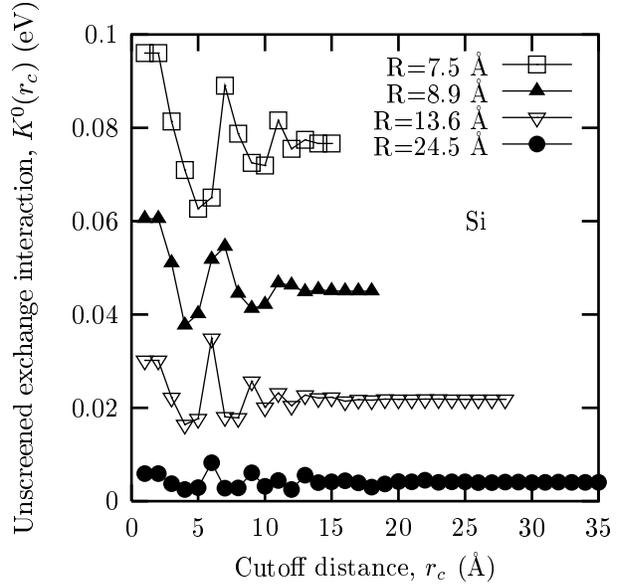
\begin{figure}
  \caption{Unscreened exchange
energy, Eq.~(\ref{eq:cutoff}), as a function of cutoff distance, 
with the Coulomb potential replaced
by a cutoff potential for various radii of Si spherical
quantum dots. The energies are for the highest hole wave function and the lowest electron
wave function. 
The curves show that there is an oscillation region for small cutoff distances
followed by a saturation region beyond 15~\AA\@.
This saturation suggests that the effective range of the exchange interaction in Si
quantum dots is around 15~\AA\@ regardless of the dot radius.}
  \label{fig:si_exch}
\end{figure}

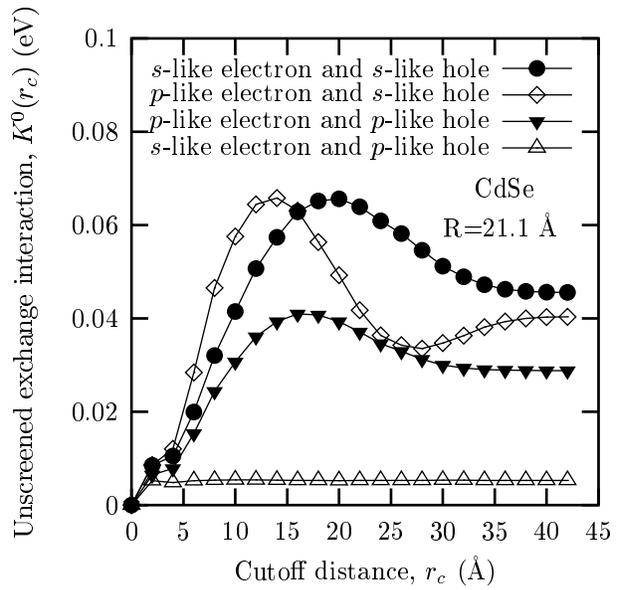
\begin{figure}
\caption{Unscreened exchange
energy, Eq.~(\ref{eq:cutoff}), as a function of cutoff distance,
with the Coulomb potential replaced by a cutoff potential for 
 the CdSe spherical quantum dot of
radius $R$=21.1~\AA\@. The unscreened exchange energy of four different
types of electron-hole configurations is shown.
The electron and hole configurations are labeled by the dominant
angular-momentum component of their envelope functions.\cite{notation_eh} 
Except for the
$s$-like electron and $p$-like hole configuration, the variation of the 
exchange interaction extends over the whole dot.}
  \label{fig:cdse_exch}
\end{figure}

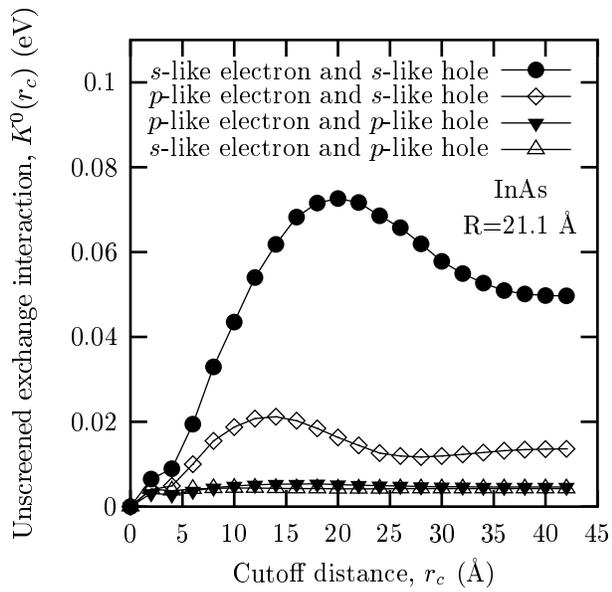
\begin{figure}
\caption{Unscreened exchange
energy, Eq.~(\ref{eq:cutoff}), as a function of cutoff distance,
with the Coulomb potential replaced by a cutoff potential for 
the InAs spherical quantum dot of
radius R=21.1~\AA\@. The unscreened exchange energy of four different
types of electron-hole configurations is shown.
 The electron and hole configurations are labeled by  the dominant
angular-momentum component of their envelope functions.\cite{notation_eh} 
Long-range
exchange interactions appear for the $s$-like hole with both the
$s$-like electron and the $p$-like electron.}
  \label{fig:inas_exch}  
\end{figure}

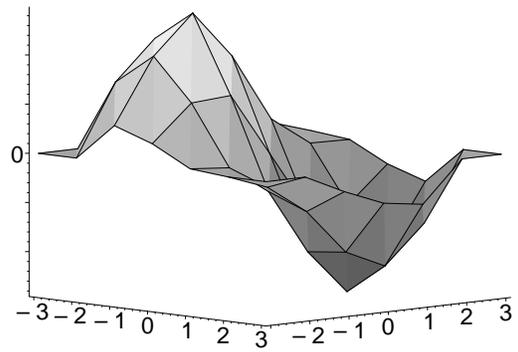
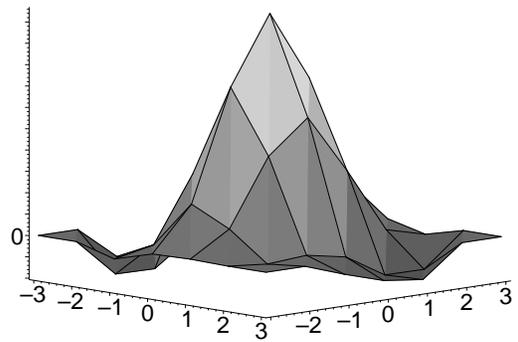
\begin{figure}
  \caption{Exchange charge density  $q(\vec R)_{eh}$ from Eq.~(\ref{eq:charge}) of (a)
the $s$-like electron and $s$-like hole configuration, and of  
(b) the $p$-like electron and $s$-like hole configuration for the CdSe quantum dot
with radius 21.1~\AA\@.
The exchange charge density is plotted in a plane through the
center of the dot. The unit of the horizontal axes is the lattice constant of CdSe.  
The plots show that the orthogonality between the electron and hole wave functions
is global not local, with a $p$-like oscillation or a 2$s$-like oscillation
respectively.   
These global oscillations of the exchange charge density 
lead to the long-range variation of the exchange interactions 
 in Fig.~\ref{fig:cdse_exch}.}
  \label{fig:cdse_charge}
\end{figure}


\begin{thebibliography}{}
\bibitem{note_cdse} Recent experiments for CdSe are based on
wurtzite CdSe, but we model CdSe as a zincblende material in this study
since the small value of the crystal field splitting in CdSe (25~meV)
 makes it reasonable to approximate this semiconductor as a zincblende material.
\bibitem{monte_ex} The Monte Carlo integration method with a biased random walk needs a non-negative density function that can be used as the probability of accepting the random walks. Since the exchange integrals involve a density function that is not positive definite, this
method is not straightforward to use.
\bibitem{stm_data} We have used the difference between the first prominent peaks
for positive and negative bias voltage to calculate the STM gaps from the
published $dI/dV$ curves. No adjustment regarding charging energy 
or polarization energy is made in the STM gaps.
\bibitem{notation_eh} To be clear about this notation, `$s$-like hole' means that the coefficients of the $p$ basis orbital has $s$ symmetry in real space over the whole dot, whereas `$p$-like hole' means that the coefficients of the $p$ basis orbital has $p$ symmetry.   
For electron states, the coefficients of the $s$ basis orbital are used to
label the wave functions. For example, 
 `$s$-like electron' means that the coefficients of the $s$ basis orbital has
 $s$ symmetry, while `$p$-like electron' has $p$ symmetry in the coefficients of
 the $s$ orbital.
\end{thebibliography}
\end{document}